\documentstyle[proc]{royal}
\input{psfig.sty}

\def\deg{\circ}
\newcommand{\dg}{\nobreak^\circ}
\newcommand{\simlt}{\mbox{$\stackrel{<}{_{\sim}}$} }
\newcommand{\simgt}{\mbox{$\stackrel{>}{_{\sim}}$} }
\newcommand{\kmspmpc}{\,\hbox{km}\,\hbox{s}^{-1}\,\hbox{Mpc}^{-1}}

\newcommand{\Omm}{\Omega_{\rm m}}
\newcommand{\Oml}{\Omega_{\Lambda}}

\def\etal{{\it et al}\etstop}

\begin{document}

\title[CMB Observations]{Observations of the cosmic microwave
background and implications for cosmology and large scale structure}

\author[A.N. Lasenby \etal ]{A.N. Lasenby, A.W. Jones \& Y. Dabrowski}

\affiliation{Astrophysics Group,\\
Cavendish Laboratory,\\
Madingley Road,\\
Cambridge,\\
CB3 0HE.
}

\maketitle

\abstract
\noindent
Observations of the Cosmic Microwave Background (CMB) are
discussed, with particular emphasis on current ground-based
experiments and on future satellite, balloon and interferometer
experiments. Observational techniques and the effects of
contaminating foregrounds are highlighted.  Recent CMB data is
used with large scale structure (LSS) data to constrain
cosmological parameters and the complementary nature of CMB, LSS
and supernova distance data is emphasized.  \endabstract

\section{Introduction}

Observations of the Cosmic Microwave Background (CMB) play a
crucial role in modern cosmology. Within a few years it is
expected that the CMB power spectrum will have been measured to an
accuracy of a few percent over a wide range of angular
scales. These observations will yield an impressive amount of
information on conditions in the early universe, and on the values
of the main cosmological parameters. The focus of this review will
be on the observations themselves, both present and future, but
will also touch on results for cosmological parameters using
current CMB data and on what can be achieved by combining the CMB
with constraints from large scale structure and supernovae.

The topics discussed include: (a) a review of what it is we wish
to measure; (b) difficulties involved in the observations, in
particular the role of contaminating foregrounds; (c) a review of
some recent experiments and results; (d) some current implications
for cosmological parameters and the tie-in with large scale
structure and supernovae and (e) brief details of future
ground-based, balloon and satellite experiments. The discussion
throughout is intended to be introductory, and therefore
complementary to the more technical presentation contained in Bond
\& Jaffe (this volume).

\section{What we wish to measure}

Ultimately, the goal of microwave background observations is to
provide accurate, high-resolution maps of the CMB sky in both
total intensity and polarization. From these we can then infer the
power spectrum and use higher order moments to distinguish between
Gaussian and non-Gaussian theories. The power spectrum is
currently the prime goal of CMB observations, since we are close
to tracing out significant features in it from which the values of
cosmological parameters can be inferred. The origin of these
features, and how they link with the cosmological parameters will now be
briefly described. (For much more detailed treatments, see Turner,
and Bond \& Jaffe (this volume).)

The inflationary theory describes an exponential expansion of
space in the very early Universe. Amplification of initial quantum
irregularities then results in a spectrum of long wavelength
perturbations on scales initially bigger than the horizon size.
Central to the theory of inflation, at least in the simplest
models, is the potential $V(\phi)$ which describes the
self-interaction of the scalar inflaton field~$\phi$. (More
general multi-field models are discussed in Lyth and Riotto
\shortcite{lyth98}.) Due to the unknown nature of this potential,
and the unknown parameters involved in the theory, inflation
cannot at the moment predict the overall amplitude of the matter
fluctuations at recombination. However, the {\em form} of the
fluctuation spectrum coming out of inflation is approximately
given by
\begin{displaymath}
|\delta_k|^2 \propto k^n,
\end{displaymath}
where $k$ is the comoving wavenumber and $n$ is the `tilt' of the
primordial spectrum. The latter is predicted to lie close to 1 (the
case $n=1$ being the Harrison-Zeldovich, or `scale-invariant'
spectrum). 

An overdensity in the early Universe does not collapse under the
effect of self-gravity until it enters the Hubble radius,
$ct$. The perturbation will continue to collapse until it reaches
the Jean's length, at which time radiation pressure will oppose
gravity and set up acoustic oscillations. Since overdensities of
the same size will pass the horizon size at the same time they
will be oscillating in phase. These acoustic oscillations occur in
both the matter field and the photon field and so will induce a
series of peaks in the photon spectrum, known as the `Doppler' or
acoustic peaks.
 
The level of the Doppler peaks in the power spectrum depends on
the number of acoustic oscillations that have taken place since
entering the horizon. For overdensities that have undergone half
an oscillation there will be a large Doppler peak (corresponding
to an angular size of $\sim 1\dg$). Other peaks occur at harmonics
of this. As the amplitude and position of the primary and
secondary peaks are intrinsically determined by the sound speed
(and hence the equation of state) and by the geometry of the
Universe, they can be used as a test of the density parameter of
baryons and dark matter, as well as other cosmological constants.

Prior to the last scattering surface the photons and matter
interact on scales smaller than the horizon size. Through
diffusion the photons will travel from high density regions to low
density regions `dragging' the electrons with them via Compton
interaction. This diffusion has the effect of damping out the
fluctuations and is more marked as the size of the fluctuation
decreases. Therefore, we expect the fluctuation spectrum and
Doppler peaks to vanish at very small angular scales. This effect
is known as Silk damping \cite{silk68}.
\begin{figure}
\centerline{\psfig{file=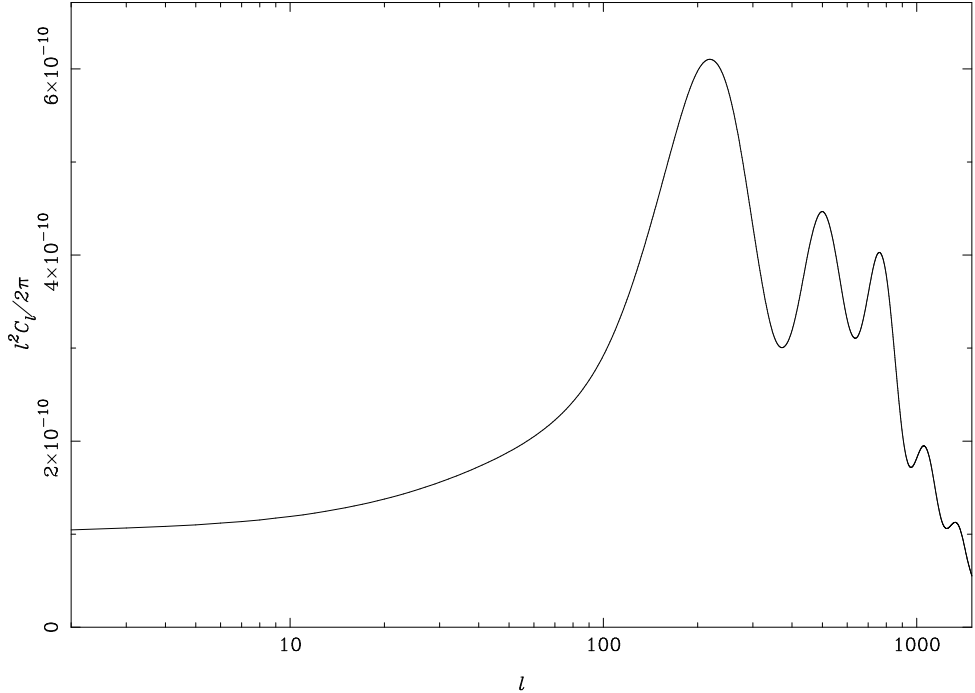,width=13cm}}
\caption{Power spectrum for standard CDM. Parameters assumed are $\Omega=1$,
$n=1$, $H_0=50\kmspmpc$ and a baryon fraction of $\Omega_b=0.04$.}
\label{fig:standard_cdm}
\end{figure}

Putting this all together, we see that on large angular scales
($\simgt 2\dg$) we expect the CMB power spectrum to reflect the
initially near scale-invariant spectrum coming out of inflation;
on intermediate angular scales we expect to see a series of peaks,
and on smaller angular scales ($\simlt 10$ arcmin) we expect to
see a sharp decline in amplitude. These expectations are borne out
in the actual calculated form of the CMB power spectrum in what is
currently the `standard model' for cosmology, namely inflation
together with cold dark matter (CDM). The spectrum for this,
assuming a density parameter, $\Omega$, of unity and standard
values for other parameters, is shown in
Figure~\ref{fig:standard_cdm}.  The quantities plotted are
$\ell(\ell+1)C_{\ell}$, versus $\ell$ where $C_{\ell}$ is defined
via
\begin{displaymath}
C_{\ell}=\langle | a_{\ell m} |^2 \rangle, \qquad \frac{\Delta T}{T}(\theta,\phi) =
\sum_{\ell,m} a_{lm} Y_{\ell m}(\theta,\phi),
\end{displaymath}
and the $Y_{\ell m}$ are standard spherical harmonics. The reason for
plotting $\ell(\ell+1)C_{\ell}$ is that it approximately 
equals the power per unit
logarithmic interval in $\ell$. Increasing $\ell$
corresponds to decreasing angular scale $\theta$, with a rough relationship
between the two of $\theta \approx 2/\ell$ radians. 
In terms of the diameter of
corresponding proto-objects imprinted in the CMB,
then a rich cluster of galaxies corresponds to a scale of about 8 arcmin, while
the angular scale corresponding to the largest scale of clustering we
know about in the
Universe today corresponds to 1/2 to 1 degree. The first large peak in the
power spectrum, at $\ell$'s near 200, and therefore angular scales near $1\dg$,
is known as the `Doppler', or `Sakharov', or `acoustic' peak. 

As stated above, the inflationary CMB power spectrum plotted in
Figure~\ref{fig:standard_cdm} is that predicted by assuming fixed
values of the cosmological parameters for a CDM model of the
Universe. In order for an
experimental measurement of the angular power spectrum to be able to
place constraints on these parameters, we must consider how the shape
of the predicted power spectrum varies in response to changes in these
parameters. In general, the detailed changes due to varying several
parameters at once can be quite complicated. However, if we restrict
our attention to the parameters $\Omega$, $H_0$ and $\Omega_b$, 
the fractional baryon density, then the situation becomes
simpler.

Perhaps most straightforward is the information contained in the
position of the first Doppler peak, and of the smaller secondary
peaks, since this is determined almost exclusively by the value of
the total $\Omega$, and varies as $\ell_{\rm peak} \propto
\Omega^{-1/2}$. (This behaviour is determined by the linear size
of the causal horizon at recombination, and the usual formula for
angular diameter distance.)  This means that if we were able to
determine the position (in a left/right sense) of this peak, and
we were confident in the underlying model assumptions, then we
could read off the value of the total density of the Universe. (In
the case where the cosmological constant, $\Lambda$, was non-zero,
we would effectively be reading off the combination $\Omega_{\rm
matter}+\Omega_{\Lambda}$.)  This would be a determination of
$\Omega$ free of all the usual problems encountered in local
determinations using velocity fields etc.

Similar remarks apply to the Hubble constant. The {\em height\/} of
the Doppler peak is controlled by a combination of $H_0$ and the
density of the Universe in baryons, $\Omega_b$. We have a constraint
on the combination $\Omega_b H_0^2$ from nucleosynthesis, and thus
using this constraint and the peak height we can determine $H_0$
within a band compatible with both nucleosynthesis and the CMB.
Alternatively, if we have the power spectrum available to good
accuracy covering the secondary peaks as well, then it is possible to
read off the values of $\Omega_{\rm tot}$, $\Omega_b$ and $H_0$
independently, without having to bring in the nucleosynthesis
information. The overall point here, is that the power spectrum of the
CMB contains a wealth of physical information, and that once we have
it to good accuracy, and have become confident that an underlying
model, such as inflation and CDM, is correct, then we can use the
spectrum to obtain the values of parameters in the model, potentially
to high accuracy. This will be discussed further below both in the
context of the current CMB data, and in the context of what we can
expect in the future.

\subsection{Polarization}

\begin{figure}
\centerline{\psfig{file=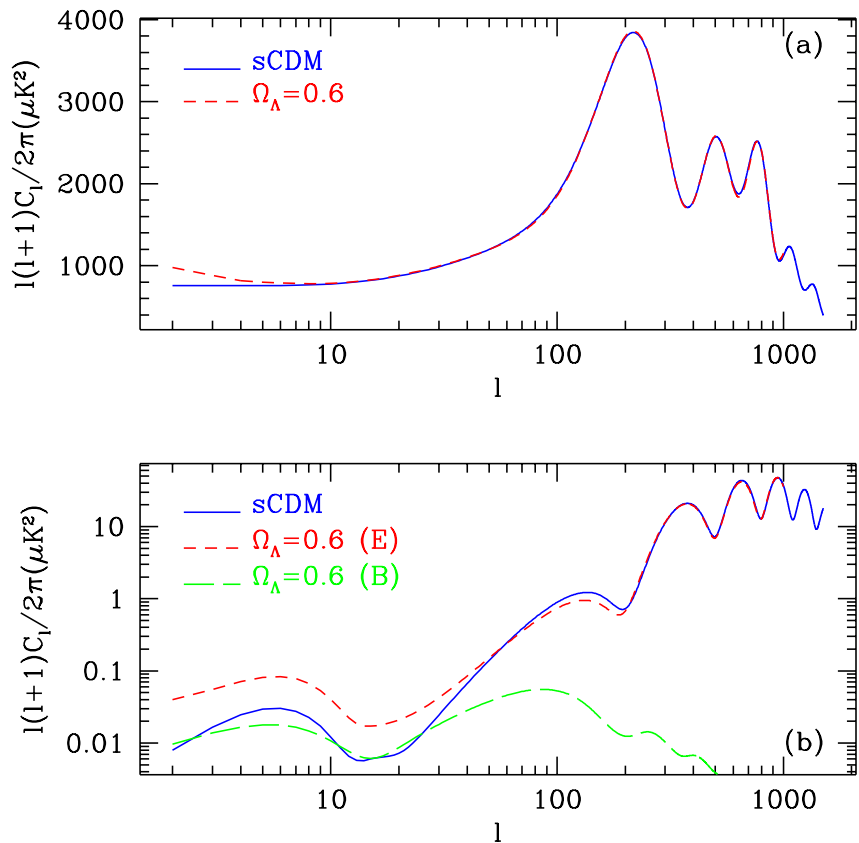,width=12cm}}
\caption{(a) Power spectrum for standard CDM shown versus a model
including a cosmological constant and non-zero tensor
component. (b) The same but for polarization ($E$ and $B$ refer to
two independent modes of the polarization). Figure
taken from Zaldarriaga \etal \protect\shortcite{zaldarriaga97}.}
\label{fig:polzn-spectrum}
\end{figure}

As well as the total intensity spectrum, we wish to measure the
polarization power spectrum, and to check for non-Gaussianity in
total intensity maps. The former will be discussed further in
Section~6d.ii in the context of a proposed new
instrument. We note here, however, that polarization information
could be very important in breaking degeneracies that occur
between parameters if just the total intensity power spectrum is
available. In the above discussion we have omitted details of the
effects of a tensor component of the fluctuation spectrum, or of
the effects of early reionization of the universe and a non-zero
cosmological constant. Different combinations of these parameters
can produce power spectra which are identical to high precision
over a large range of $\ell$ \cite{efstathiou98}. 
However, as
pointed out by Zaldarriaga \etal \shortcite{zaldarriaga97},
polarization information can break this degeneracy. This is
illustrated in the power spectra shown in
Fig.~\ref{fig:polzn-spectrum}. 
The top panel shows two total
intensity spectra, for different models, which are virtually
indistinguishable. The bottom panel is for the corresponding
polarization power spectra, and shows differences which would be
measurable by the MAP or Planck satellites (see below). Another
useful feature of the lower panel plot, is to show that the peak
of the polarization power spectra occurs at somewhat smaller
scales than for total intensity --- at $\ell$'s of around
500--1000 for the models shown here. 

\subsection{Non-Gaussianity}

As regards non-Gaussian features, as would be expected in
topological defect theories for example, this is a very large
field, a summary of which will not be attempted here. Two quick
points are worth making, however. The first is that recently
evidence has been claimed, for the first time, for non-Gaussianity
in the COBE data \cite{ferreira98}. This seems to occur only at a
particular multipole ($\ell = 16$), but is apparently highly
significant there. Secondly, an important point about the
non-Gaussian signature from cosmic strings, is that quite high
angular resolution (possibly better than 2 arcmin) may be
necessary in order to see it against the superimposed Gaussian
imprint from recombination. This has recently been emphasized by
Magueijo \& Lewin \shortcite{magueijo97}, and is of significance
for any attempt to image the Kaiser-Stebbins effect for cosmic
strings \cite{kaiser84} directly (see Section~6d.i).

\section{Experimental Problems and Solutions}

\subsection{The Contaminants}
The detection of CMB anisotropy at the level $\Delta T/T\sim
10^{-5}$ is a challenging problem and a wide range of experimental
difficulties occur when conceiving and building an experiment. We
will focus here particularly on the problem caused by
contamination by foregrounds and the solutions that have been
adopted to fight against them. The anisotropic components that are
of essential interest are (i) The Galactic dust emission which
becomes significant at high frequencies (typically $>100$~GHz);
(ii) The Galactic thermal (free-free) emission and non-thermal
(synchrotron) radiation which are significant at frequencies lower
than typically $\sim$~30~GHz; (iii) The presence of point-like
discrete sources; (iv) The dominating source of contamination for
ground- and balloon-based experiments is the atmospheric emission,
in particular at frequencies higher than $\sim$~10~GHz.

\subsection{The Solutions}
\label{sec:solns}
A natural solution is to run the experiment at a suitable
frequency so that the contaminants are kept low. There exists a
window between $\sim$~10 and $\sim$~40~GHz where both atmospheric
and Galactic emissions should be lower than the typical CMB anisotropies.
For example, the Tenerife experiments are running at $10$, $15$
and $33~{\rm GHz}$ and the Cambridge Cosmic Anisotropy Telescope
(CAT) at $15~{\rm GHz}$.  However, in order to reach the level of
accuracy needed, spectral discrimination of foregrounds using
multi-frequency data has now become necessary for all
experiments. This takes the form of either widely spaced
frequencies giving a good `lever-arm' in spectral discrimination
(e.g. Tenerife, COBE, most balloon experiments), or a closely
spaced set of frequencies which allows good accuracy in
subtraction of a particular known component (e.g. CAT, the forthcoming
VSA, and the Saskatoon experiment).

Concerning point (iv), three basic techniques, which are all still
being used, have been developed in order to fight against the
atmospheric emission problem:

The Tenerife experiments are using the {\it switched beam} method.
In this case the telescope switches rapidly between two or more
beams so that a differential measurement can be made between two
different patches of the sky, allowing one to filter out the
atmospheric variations.

A more modern and flexible version of the switched beam method is the
{\it scanned beam} method (e.g. Saskatoon and Python telescopes).
These systems have a single receiver in front of which a continuously
moving mirror allows scanning of different patches of the sky.
The motion pattern of the mirror can be re-synthesised by software.
This technique provides a great flexibility regarding the angular-scale of the
observations and the Saskatoon telescope has been very successful
in using this system to provide results on a range of angular scales.

Finally, an alternative to differential measurement is the use of
interferometric techniques. Here, the output signals from each of
the baseline horns are cross-correlated so that the Fourier
coefficients of the sky are measured. In this fashion one can very
efficiently remove the atmospheric component in order to
reconstruct a cleaned temperature map of the CMB.  The Cosmic
Anisotropy Telescope (CAT) operating in Cambridge has proved this
method to be very successful, giving great expectations for the
Very Small Array (VSA) currently being built and tested in
Cambridge (jointly with Jodrell Bank) for siting in Tenerife.
American projects such as the Cosmic Background Imager (CBI) and
the Very Compact Array (VCA/DASI) are also planning to use this
technique (see below). We now discuss further experimental points
in the context of the experiments themselves, with particular
emphasis on recent ground-based experiments from which the first
evidence for a peak in the power spectrum is emerging.

\section{Updates and Results on Various Experiments}

\subsection{The Tenerife Switched-Beam Experiments}

Due to the stability of the atmosphere and its transparency
\cite{dec40-I}, the ${\rm Iza\tilde{n}a}$ observatory of the
Tenerife island is becoming very popular for cm/mm observations of
the CMB (e.g. the Tenerife experiments, IAC-Bartol, VSA).  The
three Tenerife experiments ($10$, $15$ and $33~{\rm GHz}$) are
each composed of two horns using the switched beam technology.
The observations take advantage of the Earth rotation and consist
of scanning a band of the sky at a constant declination. The scans
have to be repeated over several days in order to achieve
sufficient accuracy.  The angular resolution is $\sim~5$ degrees,
and therefore provides a useful point on the power spectrum
diagram between COBE resolution ($\sim 10^{\deg}$) and smaller
scale experiments.  Davies \etal \shortcite{dec40-I} provides a
detailed description of the Tenerife experiments.
 
The first detection at Tenerife (Dec+$40^{\deg}$), which dates
back to 1994~\cite{hannat94,dec40-paperII}, clearly reveals common
structures between the three independent scans at $10$, $15$ and
$33~{\rm GHz}$.  The consistency between the three channels gave
confidence that, for the first time, identifiable individual
features in the CMB were detected~\cite{anl-capri}.  Subsequently
this was confirmed by comparing directly to the COBE DMR
data~\cite{line95,sh-capri}.

Bunn, Hoffman \& Silk \shortcite{bunn96} have applied a Wiener
filter to the COBE DMR data in order to perform a prediction for
the Tenerife experiments over the region $35^{\deg}<{\rm
Dec}<45^{\deg}$.  Assuming a CDM model, the COBE angular
resolution was improved using the Wiener filtering in order to
match the Tenerife experiments' resolution.  The prediction has
been observationally verified~\cite{carlos-dec35}, giving great
confidence that the revealed features are indeed tracing out the
seed structures present in the early universe.

\subsubsection{Latest results}

There is now enough data to perform a 2-D sky reconstruction
\cite{aled-10ghz} for the $10~{\rm GHz}$ and $15~{\rm GHz}$
experiments ($33~{\rm GHz}$ to follow shortly).  Eight separate
declination scans have been performed over the full range in RA
from Dec+$27.5^{\deg}$ up to Dec+$45^{\deg}$ in steps of
$2.5^{\deg}$.  This allows the reconstruction with reasonable
accuracy of a strip in the sky of $90^{\deg}\times 17.5^{\deg}$ in
an area away from major point sources and the Galactic plane.  An
important aspect in obtaining accurate results is, first of all,
to allow for atmospheric correlations between the different scans
\cite{gutierrez97}.  Secondly, and probably more importantly, to
be aware that the maps are sensitive to unresolved discrete radio
sources (typically at the Jy level in the Tenerife field) in
addition to the CMB. Special analysis has been performed in order
to remove these sources which have to be monitored continuously
since they are variable on the time scales involved.  This
monitoring task is done in collaboration with M. and H. Aller
(Michigan) who have a data-bank of information on these sources.

The 10~GHz 2-D map is likely to include a significant Galactic
contribution; however it is believed that this contribution is
much smaller for the 15~GHz map which reveals intrinsic CMB
anisotropies on a 5 degrees scale. Likelihood analysis on the
reconstructed 15~GHz 2-D map is in preparation and will be
published shortly. Previous results~\cite{dec40-paperII} are:
$\delta T=\left[l(l+1)C_l/\left(2\pi\right)\right]^{1/2}=
34^{+15}_{-9}~\mu{\rm K}$ at $l\sim 18$ (see Table~\ref{tab:1} and
Figure~\ref{fig:cmb_iras_data}).

One of the next steps concerning the data analysis is to use the
Maximum Entropy Method for frequency separation on the spherical
sky, in conjunction with all sky maps such as the Haslam
408~MHz~\cite{haslam82}, IRAS, Jodrell Bank (5~GHz) and COBE.  The
resulting frequency information will allow much improved
separation of the synchrotron, free-free, dust and CMB
components, which is an exciting prospect.

\subsection{IAC-Bartol}

This experiment runs with four individual channels (91, 142, 230
and 272~GHz) and is also located in Tenerife where the dry
atmosphere is required for such high frequencies.  This novel
system uses bolometers which are coupled to a 45~cm diameter
telescope. The angular resolution is approximately $2^{\deg}$ (see
elsewhere~\cite{piccirillo91,piccirillo93,piccirillo97} for
instrument details and preliminary results).

This switched beam system has performed observations at constant
declination (Dec+$40^{\deg}$), overlapping one of the drift scans
of the Tenerife experiments.  Atmospheric correlation techniques
between the different frequency channels have been applied in
order to remove the strong atmospheric component present in the
three lowest channels~\cite{femenia98}.  The Galactic synchrotron
and free-free emissions are likely to be much smaller than the CMB
fluctuations at these frequencies. On the other hand, the Galactic
dust emission has been corrected using DIRBE and COBE DMR maps.
Finally, the contamination by point-like sources was removed by
multi-frequency analysis on known and unknown sources. The results
obtained are $\delta T=113^{+66}_{-60}~\mu{\rm K}$ at $l\sim 33$
and $\delta T=55^{+27}_{-22}~\mu{\rm K}$ at $l\sim 53$ (see
Table~\ref{tab:1}. One can notice (e.g. by comparison with the
expected curve in Fig.~\ref{fig:cmb_iras_data}) that the $l\sim
33$ point is well off the expected value, however, tests show that
the atmospheric component is still very high in this $\delta T$
value. The $l\sim 53$ point seems to be in better agreement with
results from the Saskatoon or Python experiments.

\begin{table}
\caption{Some Current Ground Based Experiments}
\footnotesize
\vspace{0.4cm}
\begin{center}
\begin{tabular}{c c c c c c}
\hline
Experiment & Frequency & Angular Scale & Site/Type & $l$ & $\delta T$ \\
\hline & & & & &\\
Tenerife & 10, 15, 33~GHz& $\sim 5^{\deg}$ & Tenerife (Switched Beam) &
    $18^{+9}_{-7}$ &$ 34^{+15}_{-9}$ \\ & & & & &\\

IAC-Bartol & 91, 142, & $\sim 2^{\deg}$ & Tenerife &
    $33^{+24}_{-13}$ & $113^{+66}_{-60}$ \\ 
    & 230 and 272~GHz &  & (Switched beam) &
    $53^{+22}_{-13}$ & $55^{+27}_{-22}$ \\ & & & & &\\

Python III& 90~GHz & $0.75^{\deg}$ & South Pole &
    $87^{+18}_{-38}$ & $60^{+15}_{-13}$ \\
    & & & (Scanned Beam) & $170^{+69}_{-50}$ & $66^{+17}_{-16}$ \\

Python I, II \& III& & & & $139^{+99}_{-34}$ & $63^{+15}_{-14}$ \\ & & & & &\\

%    & & & & $87^{+39}_{-29}$ & $52^{+8}_{-5}$ \\ 
%    & 6/12 channels & $0.5^{\deg}-3^{\deg}$ & Canada &
%    $166^{+30}_{-43}$ & $74^{+7}_{-6}$ \\
%Saskatoon & between & & (Scanned Beam) & $237^{+29}_{-41}$ & $91^{+10}_{-8}$\\
%    & 26 and 46~GHz  &  & & $286^{+24}_{-38}$ & $92^{+12}_{-10}$ \\
%    & & & & $349^{+44}_{-41}$ & $74^{+19}_{-28}$ \\ & & & & &\\

    & & & & $87^{+39}_{-29}$ & $49^{+8}_{-5}$ \\ 
    & 6/12 channels & $0.5^{\deg}-3^{\deg}$ & Canada &
    $166^{+30}_{-43}$ & $69^{+7}_{-6}$ \\
Saskatoon & between & & (Scanned Beam) & $237^{+29}_{-41}$ & $85^{+10}_{-8}$\\
    & 26 and 46~GHz  &  & & $286^{+24}_{-38}$ & $86^{+12}_{-10}$ \\
    & & & & $349^{+44}_{-41}$ & $69^{+19}_{-28}$ \\ & & & & &\\

OVRO & 14.56 and 32 GHz & $\sim 0.1^{\deg}-0.4^{\deg}$ &
    Owens Valley (Switched) & $589^{+167}_{-228}$ & $56^{+8.5}_{-6.6}$ \\ & & & & &\\

CAT & 13 to 17~GHz & $0.5^{\deg}$ & Cambridge, UK &
    $615^{+110}_{-60}$ & $55^{+11}_{-11}$\\
    & & & Interferometer (3) & $422^{+90}_{-50}$ & $57^{+11}_{-14}$\\
    & & & & &\\
 \hline
\label{tab:1}
\end{tabular}
\end{center}
\end{table}

\subsection{Python}

This experiment is using a single bolometer mounted on a 75cm
telescope and operating at the single frequency of 90GHz with a
$0.75^{\deg}$ FWHM beam.
Python is located at the Amundsen-Scott South Pole Station
in Antarctica. It is performing extremely well in terms of mapping rather
large regions of the sky (currently $22^{\deg}\times 5.5^{\deg}$).
Three seasons of observations have been analysed so far
(Python~I~\cite{dragovan94}, Python~II~\cite{ruhl95} and
Python~III~\cite{platt97}). In addition to the power-spectrum
results of Python~III (see Table~\ref{tab:1} and Figure~\ref{fig:cmb_iras_data}),
the combined analysis of Python~I, II \& III gives an estimate of the
power-spectrum angular spectral index~\cite{platt97}: $m=0.16^{+0.2}_{-0.18}$
which is consistent with a flat-band power model (i.e. $m=0$).

A point where the Python experiment differs from all the others is
its single frequency measurement.  All the experiments discussed
here are using either widely spaced frequencies (e.g. Tenerife
experiments, COBE) or closely patched bands of different
frequencies (e.g. the interferometers discussed in
Section~6a).  As mentioned above, multi-frequency
analysis allows identification and correction of the
contaminating component. However, near the pole, the atmospheric
emission is believed to be small, while at 90~GHz the Galactic
dust contribution is estimated to be as small as $\sim 2\mu K$.
On the other hand, 17 known point-like sources are present in the
Python field, which are estimated to give a 2\% effect in the
final result.  The brightest source may contribute up to $50~\mu
K$ in a single beam and ideally source removal using information
from a separate telescope at the same frequency is required.

Python~IV and V data have already been taken and the analysis
should provide power-spectrum estimations very shortly; see Kovac
\etal \shortcite{kovac97} and Coble \etal \shortcite{coble98} for
details about the ${\rm IV^{th}}$ and ${\rm V^{th}}$ seasons
respectively.

\subsection{Saskatoon current status}

The Saskatoon experiment is a scanned beam system which operates
with 6 or 12 independent channels at frequencies between 26 and
46~GHz. The observations cover the North Celestial Pole
with angular scales from $0.5^{\deg}$ to $3^{\deg}$.
The experiment has been running from 1993 to 1995 
and details of the instrument as well as early results
can be found elsewhere \cite{wollack93,wollack96}. To find more 
details about the data analysis and recent results, see for example
Wollack \etal \shortcite{wollack97},
Netterfield \etal \shortcite{netterfield97} and
Tegmark \etal \shortcite{tegmark97}.

The 5 Saskatoon results (see Table~\ref{tab:1}) are crucial
in constraining the position of the first Doppler peak
(see Figure~\ref{fig:cmb_iras_data}) and therefore the cosmological parameters.
The overall flux calibration of the Saskatoon data was
known to have a $\pm$14\% error,
affecting significantly estimates of Hubble's parameter ($H_0$)
for spatially flat models
for example.
However, recent work from Leitch~{\it et al}. (private communication)
who carried out joint observations of Cassiopea~A and Jupiter,
allows the reduction of this uncertainty. The latest
calibration is now known with an estimated error of $\sim 4$\%.
%(note that results from
%Netterfield~{\it et al}.~\cite{netterfield97} have to be increased
%by 7\%,
%see Table~\ref{tab:1} and Figure~\ref{fig:ps}).

Recent work on the foreground analysis of the Saskatoon field has
been carried out by Oliveira-Costa \etal \shortcite{oliveira97}.
These authors found no significant contamination by point-like sources.
However, they report a marginal correlation between the DIRBE
$100 \, \mu {\rm m}$ and Saskatoon $Q$-Band maps which is likely to be
caused by Galactic free-free emission.
This contamination is estimated to cause previous CMB results in
this field to be
over-estimated by a factor of 1.02.

\subsection{Mobile Anisotropy Telescope}

The Mobile Anisotropy Telescope (MAT) is using the same optics and
technology as Saskatoon at a high-altitude site in Chile (Atacama
plateau at 5200m).  This site is believed to be one of the best
sites in the world for millimetre measurements and is now becoming
popular for other experiments (e.g. the Cosmic Background
Interferometer, see Section~6a below) because of its dry weather.
The experiment is mounted on a mobile trailer which will be towed
up to the plateau for observations and maintenance.  The relevant
point where MAT differs from the Saskatoon experiment is the
presence of an extra channel operating at $140~{\rm GHz}$.  This
will greatly improve the resolution and should provide results
well over the first Doppler peak. Data has already been taken over
the last few months at $140~{\rm GHz}$ and is currently being
analysed.  See the MAT www-page~\cite{herbig98} for a full
description of the project.

\subsection{The Cosmic Anisotropy Telescope}
\label{cat}

The Cosmic Anisotropy Telescope (CAT) is a three element,
ground-based interferometer telescope, of novel
design~\cite{cat93}.  Horn-reflector antennas mounted on a
rotating turntable, track the sky, providing maps at four
(non-simultaneous) frequencies of 13.5, 14.5, 15.5 and
16.5~GHz. The interferometric technique ensures high sensitivity
to CMB fluctuations on scales of $0.5^{\circ}$, (baselines $\sim
1$m) whilst providing an excellent level of rejection to
atmospheric fluctuations. Despite being located at a relatively
poor observing site in Cambridge, the data is receiver noise
limited for about 60\% of the time, proving the effectiveness of
the interferometer strategy.  The first observations were
concentrated on a blank field (called the CAT1 field), centred on
RA $08^h$ $20^m$, Dec.  $+68^{\circ}$ $59'$, selected from the
Green Bank 5~GHz surveys under the constraints of minimal discrete
source contamination and low Galactic foreground. The data from
the CAT1 field were presented in O'Sullivan \etal
\shortcite{osull95} 
and Scott \etal \shortcite{cat-paper96}.

\begin{figure}
\centerline{\psfig{file=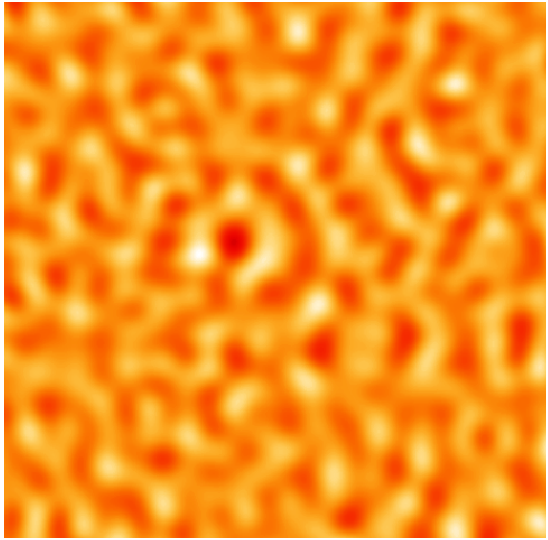,height=2.8in}}
\caption{16.5~GHz CAT image of $6\dg \times 6\dg$ area centred on the 
CAT2 field, after discrete sources have been subtracted.
Excess power can be seen in the central $2\dg \times 2\dg$ primary 
beam (because the sensitivity drops sharply outside this area,
the outer regions are a good indicator of the noise level on the
map). The flux density range scale spans $\pm 40$~mJy per
beam.}
\label{fig:CAT-maps}
\end{figure}

Recently observations of a new blank field (called the CAT2
field), centred on RA $17^h$ $00^m$, Dec. $+64^{\circ}$ $30'$,
have been taken.  Accurate information on the point source
contribution to the CAT2 field maps, which contain sources at much
lower levels, has been obtained by surveying the fields with the
Ryle Telescope at Cambridge, and the multi-frequency nature of the
CAT data can be used to separate the remaining CMB and Galactic
components. Some preliminary results from CAT2 have been presented
in Baker \shortcite{baker97} and a more detailed paper has
recently been submitted. The 16.5~GHz map is shown in
Figure~\ref{fig:CAT-maps}.  Clear structure is visible in the
central region of this map, and is thought to be actual structure,
on scales of about $1/4\dg$, in the surface of last scattering.
When interpreting this map, however, it should remembered that
for an interferometer with just three horns, the `synthesised' beam
of the telescope has large sidelobes, and it is these sidelobes that
cause the regular features
seen in the map. In the full analysis of the data, these
sidelobes must be carefully taken into account. 

\begin{figure}[t]
\centerline{\psfig{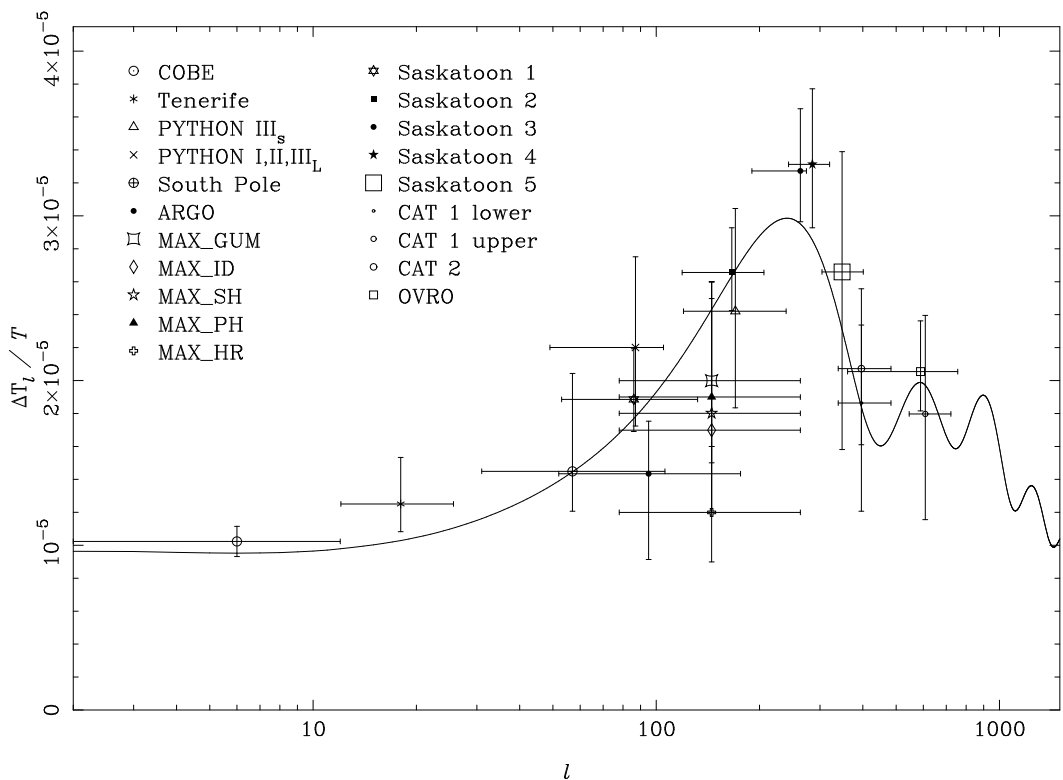}}
\caption{CMB data points and the best fit power
spectrum from fitting jointly with IRAS 1.2 Jy data, taken from
Webster \etal \protect\shortcite{cmb-iras}. 
The COBE data point is the 4-year COBE DMR result from
Bennett \etal \protect\shortcite{bennett96}.} 
\label{fig:cmb_iras_data}
\end{figure}

For an interferometer, `visibility space' correlates directly with the space
of spherical harmonic coefficients $\ell$ discussed earlier, and the data
may be used to place constraints directly on the CMB power spectrum in two
independent
bins in $\ell$. These constraints, along with those from the other experiments, 
are shown in Figure~\ref{fig:cmb_iras_data}.

\subsection{Owens Valley Radio Observatory}

The Owens Valley Radio Observatory (OVRO) telescopes have been used since
1993 for observation of the CMB at $14.5~{\rm GHz}$ and $32~{\rm GHz}$.
The RING40M experiment uses the 40-meter telescope ($14.5~{\rm GHz}$ channel)
while the RING5M experiment is mounted on the 5.5-meter telescope
($32~{\rm GHz}$ channel). Both experiments have an angular resolution 
of $\sim 0.12^{\deg}$. Details about these experiments can be found
in Readhead \etal \shortcite{readhead89} or
in Myers \etal \shortcite{myers93} for example.

36 fields at Dec+$88^{\deg}$, each separated by 22 arcmin, have
been observed around the North Galactic Pole. Using these data,
Leitch \etal \shortcite{leitch97} report an anomalous component of
Galactic emission. Further work on the same data which has just
become available \cite{leitch98}, gives the following estimate for
the CMB component: $\delta T=56^{+8.5}_{-6.6}~\mu \, {\rm K}$ at
$l\sim 589$.  As seen in Figure~\ref{fig:cmb_iras_data}, this new
OVRO result seems to agree well with the CAT estimations and
therefore helps in constraining the position of the first Doppler
peak.

\section{Using CMB, Large Scale Structure and Supernovae data to
constrain cosmological parameters}

\label{sec:like}
As mentioned earlier, by comparing the observed CMB power spectrum
with predictions from cosmological models one can estimate
cosmological parameters. This has become an area of great current
interest, with many groups carrying out the analyses for a range
of assumed models.  \cite{dec40-paperII,omega-paper,lineweaver97}
(see also Bond \& Jaffe, this volume). Generally speaking, the
results of using CMB data alone to do this are broadly consistent
with the expected range of cosmological parameters, though perhaps
with a tendency for $H_0$ to come out rather low (assuming
spatially flat models). In an
independent manner, similar predictions can be achieved by
comparing Large Scale Structure (LSS) surveys with cosmological
models~\cite{willick97,fisher96,heavens95}.  Recently, Webster
\etal \shortcite{cmb-iras} have used full likelihood calculations
within a specific model in order to join together CMB and LSS
predictions. This approach is complementary to that of Gawiser \&
Silk \shortcite{gawiser98}, who used a compilation of large scale
structure and CMB data to assess the goodness of fit of a wide
variety of cosmological models. Webster \etal use results from various
independent CMB experiments (the compilation used is that shown in
Fig.~\ref{fig:cmb_iras_data}) together with the IRAS 1.2~Jy galaxy
redshift survey and parametrise a set of spatially flat models.
Because the CMB and LSS predictions are degenerate with respect to
different parameters (roughly: $\Omega_m {\rm
vs}~\Omega_{\Lambda}$ for CMB; $H_0$ and $\Omega_m {\rm vs}~b_{\rm
iras}$ for LSS), the combined data likelihood analysis allows the
authors to break these degeneracies, giving new parameter
constraints. Note $\Omega_m$ is the overall {\it matter}
density, satisfying $\Omm+\Oml=1$.

\begin{figure}[t]
\centerline{\psfig{file=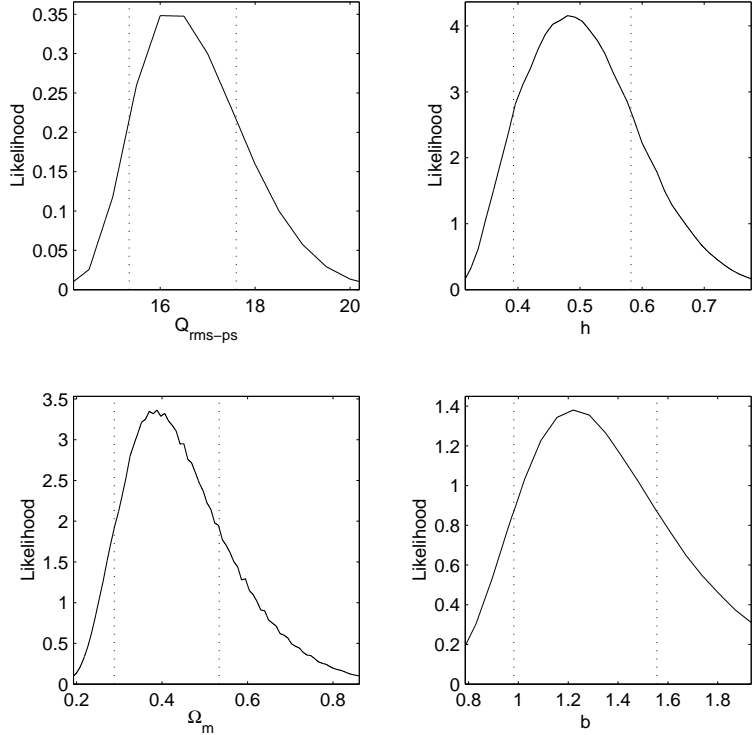,width=10cm}}
\caption{The one-dimensional marginalised probability
distributions for each of the four parameters: $Q_{\rm rms-ps}$,
the power spectrum normalization; $h=H_0/100~{\rm km~s^{-1}}$;
$\Omm$ and $b_{\rm iras}$, the (assumed linear) biasing level for
IRAS galaxies. The vertical dashed lines denote the 68\%
confidence limits. The horizontal plot limits are at the 99\%
confidence limits. (Figure taken from Webster \etal \protect\shortcite{cmb-iras}.)}
\label{fig:margplots}
\end{figure}

The results of the joint analysis are given here, as being
indicative of the current constraints available from the CMB
data. Fig.~\ref{fig:margplots} shows the final 1-dimensional
probability distributions for the main cosmological parameters
after marginalizing over each of the others. The constraint
$\Omega_b h^2 = 0.024$, where $h=H_0/100~{\rm km~s^{-1}}$, has
been assumed, close to the value expected from primordial
nucleosynthesis.

The best fit results from the joint analysis of the two data sets
on all the free parameters are shown in
Table~\ref{tab:best-fit-values}. A detailed discussion of these
estimates and comparison with other results
is contained in Webster \etal \shortcite{cmb-iras}, but in broad
terms it is clear that fairly sensible values have resulted, which
is encouraging for future prospects within this area.
\begin{table}
\centering
\begin{tabular}{@{}lc r@{.}l @{$\,<\,$} c @{$\,<\,$} r@{.}l}
             &Free parameters           \\
$\Omm$    &$0.39$    &  $0$ & $29$ &$\Omm$&$0$&$53$\\
$h$          &$0.53$    &  $0$ & $39$ &$h$      &$0$&$58$ \\
$Q$ ($\mu$K) &$16.95$   &  $15$& $34$ &$Q$      &$17$&$60$\\
$b_{\rm iras}$        &$1.21$    &  $0$ & $98$ &$b_{\rm iras}$    &$1$&$56$\\
\\
             &Derived  parameters \\
$\Omega_b$   &$0.085$\\
$\sigma_8$        &$0.67$\\
$\sigma_{8,{\rm iras}}$        &$0.81$\\
$\Gamma$     &$0.15$\\
$\beta_{\rm iras}$     &$0.47$\\
Age (Gyr)    &$16.5$\\
\\
\end{tabular}
\caption{Parameter values at the joint optimum.
For the free parameters the  
68\% confidence limits are shown,
calculated for each parameter by marginalising the likelihood over the other 
variables. (Table taken from Webster \etal \protect\shortcite{cmb-iras}.) }
\label{tab:best-fit-values}
\end{table}
For a spatially flat
model, the age of the universe is given by:
\begin{displaymath}
t = \frac{2}{3 H_0} \frac{\tanh^{-1}\sqrt{\Oml}}{\sqrt{\Oml}}
\end{displaymath}
which evaluates to 16.5 Gyr in the current case, again compatible
with previous estimates.

\subsection{Combining CMB and Supernovae data}

There has recently been great interest in combining Type Ia
supernovae (SN) data with results from the CMB (e.g. Tegmark \etal
\shortcite{tegmark98}, Lineweaver \shortcite{lineweaver98}). It is
instructive to see how the {\em complementarity\/} between the
supernovae and CMB data comes about.  The key quantity for this
discussion from both the CMB and SN points of view is $R_0
S(\chi)$, which occurs in the definitions of {\em Luminosity
Distance:}
\begin{displaymath}
d_L = R_0 S(\chi) (1+z),
\end{displaymath}
and {\em Angular Diameter Distance:} 
\begin{displaymath}
d_{\theta} = R_0 S(\chi)/(1+z)
\end{displaymath}
Here $\chi$ is a comoving coordinate, and $S(\chi)$ is
$\sinh(\chi)$, $\chi$ or $\sin(\chi)$ depending on whether the universe
is open, flat or closed respectively.
For a general Friedmann-Lemaitre model, one finds that
\begin{displaymath}
R_0 S(\chi) \propto \frac{1}{|\Omega_{\rm k}|^{1/2}} {\rm sin(h)} \left\{ 
|\Omega_{\rm k}|^{1/2} \int_0^z \frac{dz'}{H(z')} \right\}
\end{displaymath}
where
\begin{eqnarray*}
\Omega_{\rm k} & = & 1 - (\Omm + \Oml), \\
H^2(z) & = & H_0^2 \left( (1+\Omm z)(1+z)^2 - \Oml z (2+z) \right).
\end{eqnarray*}
For small $z$, it is easy to show that
\begin{displaymath}
d_L \propto z + \frac{1}{2}(1-2q_0)z^2,
\end{displaymath}
where $q_0 = \frac{1}{2}(\Omm - 2\Oml)$ is the usual
deceleration parameter.

Therefore, for small $z$, SN results are degenerate along a line
of constant $q_0$.  However, the contours of equal $R_0 S(\chi)$
shift around as $z$ increases and for $z \simgt 100$, the contours
are approximately orthogonal to those corresponding to $q_0$
constant.  This is the essence of why CMB and SN results are
ideally complementary. The current microwave background data is
mainly significant in delimiting the left/right position of the
first Doppler peak in the power spectrum, and this depends on the
cosmology via the angular diameter distance formula, evaluated at
$z \sim 1000$.  Thus the CMB results will tend to be degenerate
along lines roughly perpendicular to those for the supernovae in
the $(\Omm,\Oml)$ plane. Detailed likelihood calculations using
both supernovae and CMB data are currently being carried out by
Efstathiou \etal, and should be submitted shortly.
\normalsize{%
\begin{figure}[t]
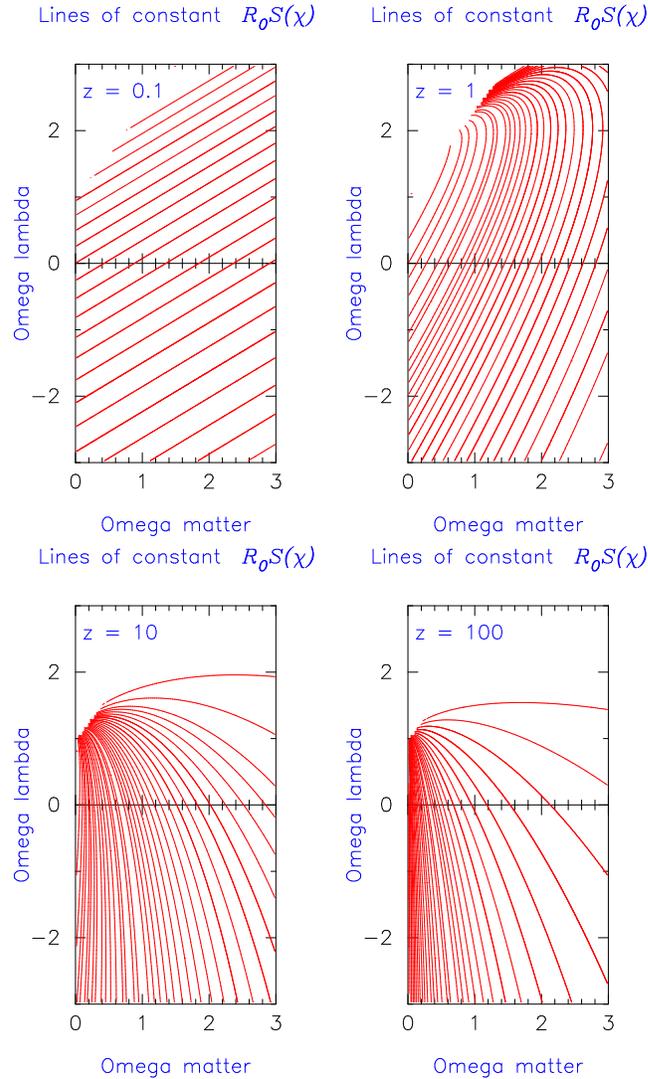

\centerline{%
\begin{tabular}{cc}
\mbox{\psfig{figure=santander_plot_z_0p1.ps,angle=-90,width=4cm}} 
& 
\mbox{\psfig{figure=santander_plot_z_1.ps,angle=-90,width=4cm}} \\
\mbox{\psfig{figure=santander_plot_z_10.ps,angle=-90,width=4cm}} 
& 
\mbox{\psfig{figure=santander_plot_z_100.ps,angle=-90,width=4cm}} 
\end{tabular}
}
\caption{Plots showing contours of constant $R_0 S(\chi)$ at
redshifts of 0.1, 1, 10 and 100.}
\end{figure}
}

\section{Future Experiments}

\subsection{Ground Based Interferometers}
\label{sec:int}

\begin{table}
\caption{Future Ground Based Experiments}
\footnotesize
\vspace{0.4cm}
\begin{center}
\begin{tabular}{c c c c c}
\hline
Experiment & Frequency & Angular Scale & Site/Type & Date \\ 
\hline
 & & & &\\

%MAT  & Saskatoon + 140~GHz & $<0.5^{\deg}$ &
%    Chile, 5200m (Scanned Beam) & 1999\\ & & & &\\
VSA  & 26 to 36~GHz & $0.25^{\deg}-2^{\deg}$ &
    Tenerife (14 element interferometer)& 1999\\
% & & & &\\
CBI  & 26 to 36~GHz & $0.07^{\deg}-0.3^{\deg}$ &
    Chile (13 element interferometer)&1999 \\
% & & & &\\
DASI & 26 to 36~GHz & $0.25^{\deg}-1.4^{\deg}$ &
    South Pole (13 element interferometer)&1999 \\
 & & & &\\ 
\hline
\label{tab:2}
\end{tabular}
\end{center}
\end{table}

As seen in Section~3b,
interferometers allow accurate removal of
the atmospheric component.
Therefore special ground sites are not
always necessary in order to perform sensitive measurements, as
already seen for the 3 element Cosmic Anisotropy Telescope
(CAT) currently operating in Cambridge, UK (see above).

\begin{figure}[t]
\centerline{\psfig{file=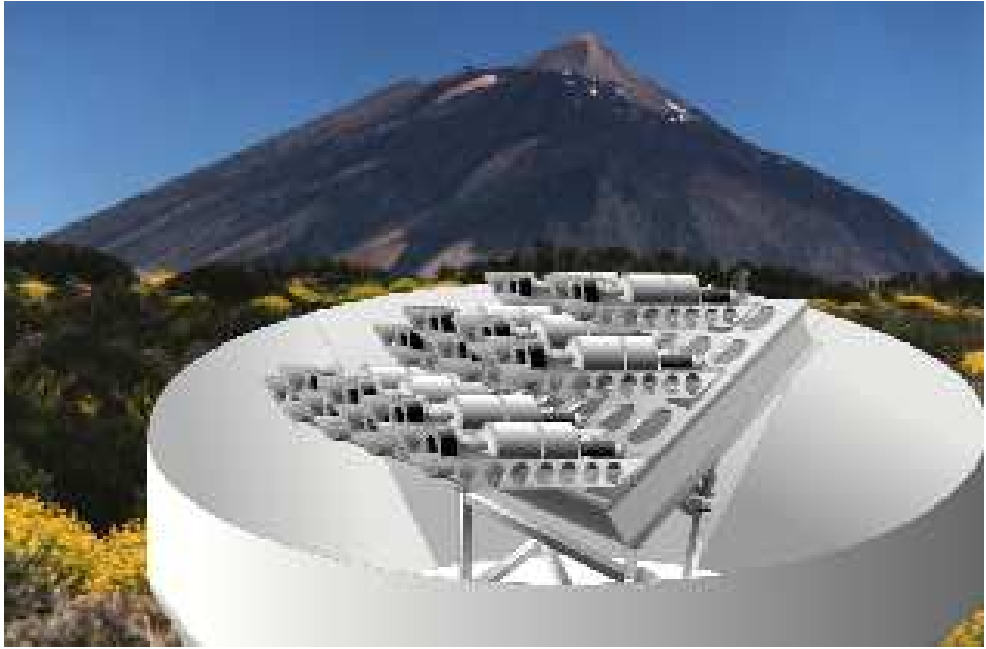,angle=-90,width=13cm}}
\caption{Artist's impression of the VSA in operation near
Mt. Teide in Tenerife.}
\label{fig:vsa-pic}
\end{figure}

The Very Small Array (VSA) is currently being built and tested in
Cambridge for siting in Tenerife and should be observing in late
1999. The 14 elements of the interferometer will operate from 26
to 36~GHz and cover angular scales from $0.25^{\deg}$ to
$2^{\deg}$ (see Table~\ref{tab:2}).  The results will consist of 9
independent bins regularly spaced from $l\sim 150$ to $l\sim 900$
on the Power Spectrum diagram~\cite{jones97}.  This will give
significant information on the second Doppler peak, while (subject
to constraints on one of $\Omega_{\rm k}$ or $\Oml$) the first
peak will be constrained accurately enough to estimate the total
density $\Omega$ and Hubble's constant $H_0$, with a 10\% error,
by the end of the year 2000. An artists impression of the VSA is
shown in Fig.~\ref{fig:vsa-pic}.
There are two other interferometer projects which will complement the
work done with the VSA:
The Cosmic Background Imager (CBI), to be operated from Chile
by a CalTech team~\cite{pearson98} and
The Degree Angular Scale Interferometer~\cite{white97,stark98} (DASI)
-- formerly Very Compact Array (VCA) -- which will be operated at the
South Pole (University of Chicago \& CARA).
They both share the same design (13 element interferometers)
and the same 
correlator operating from 26 to 36~GHz (see Table~\ref{tab:2}).
However the size of the baselines differ between CBI and DASI so that
CBI will cover angular scales from 4 to 20~arcmin while DASI will
cover the range between 15~arcmin and $1.4^{\deg}$ (similar to
the VSA).

All three of these interferometric experiments (VSA, CBI and DASI)
should be in operation by the end of 1999.

\subsection{Forthcoming balloon experiments}

The paper by Bond \& Jaffe (this volume) provides details of the
expected parameter-estimation performance of several upcoming
balloon experiments. Here we comment briefly on the
experimental aspects of the new generation of balloons.

Experiments such as MAX (e.g. Tanaka \etal \shortcite{tanaka96})
and MSAM (e.g. Cheng \etal \shortcite{cheng97})  have been very
significant in providing CMB anisotropy data points at mm and
sub-mm frequencies, and figure prominently in compilations of
current data. In order to reduce the quite large scatter
associated with the balloon data, however, the key requirement has
been to increase the effective observing time from the 9 to 12
hours of a typical launch, so that larger sky
areas can be surveyed (reducing sample variance) to greater depth.
Two main ways have evolved to achieve this. The first is the use
of array receivers. Here, instead of one pixel on the sky at each
frequency, many are used, speeding up throughput. The MAXIMA
experiment, which has grown out of the MAX program, 
has 8 simultaneous 12 arcmin pixels available, and has recently
completed its first flight. Data from this is currently being analysed.

The second method is to directly increase the time for which the
balloon can take data. This is being achieved by launching the
balloons in Antarctica and relying on the circulating winds near
the South Pole to sweep the balloon around in a roughly circular
orbit back to the launch site, where it can be recovered. Three
groups have now got funding for such experiments. These are
(a) BOOMERANG (Caltech and Berkeley) in LD (Long Duration) mode, which
will circle the pole in 7 to 14 days. (A preliminary flight of
BOOMERANG in North America has already been completed and data is
expected from this soon); (b) ACE (collaboration between UCSB,
Milan and Bologna), which unlike other balloon systems uses
heterodyne rather than bolometer technology (and correspondingly lower
frequencies). This experiment is planned to ultimately evolve to an
ultralong duration system called BEAST, which would circle the
pole in about 100 days; (c) TOPHAT (collaboration between GSFC,
Chicago and Barthol), which is also long duration but
distinguished from BOOMERANG in that the payload is mounted {\em
on top of\/} the balloon in an attempt to provide a more
systematic-free environment. Further details of all these missions
can be found in the web pages cited in the references of Bond \&
Jaffe (this volume).

\subsection{Satellite experiments}

\begin{figure}[t]
\centerline{\psfig{file=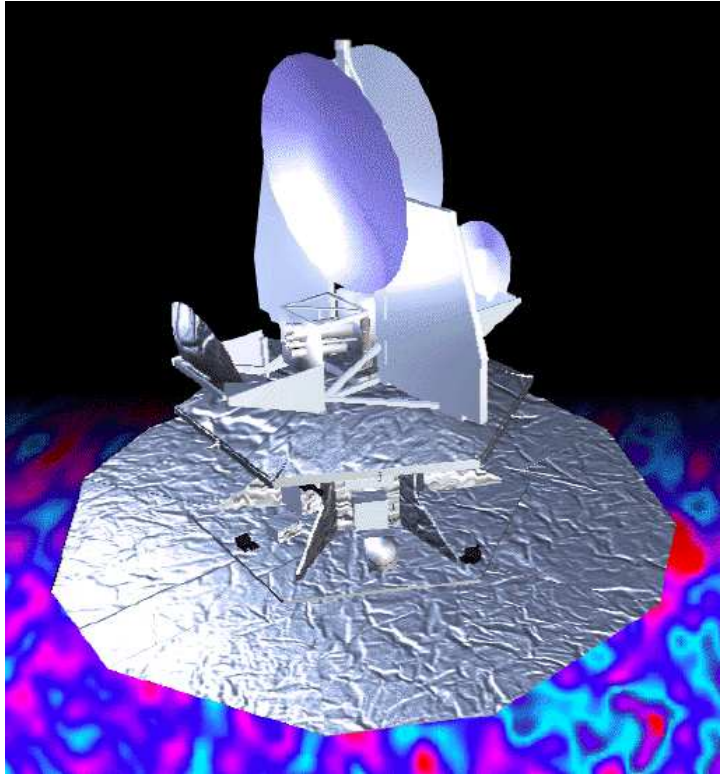,width=9.5cm}}
\caption{Artist's impression of the MAP Satellite.}
\label{fig:MAP-pic}
\end{figure}

\begin{figure}[t]
\centerline{\psfig{file=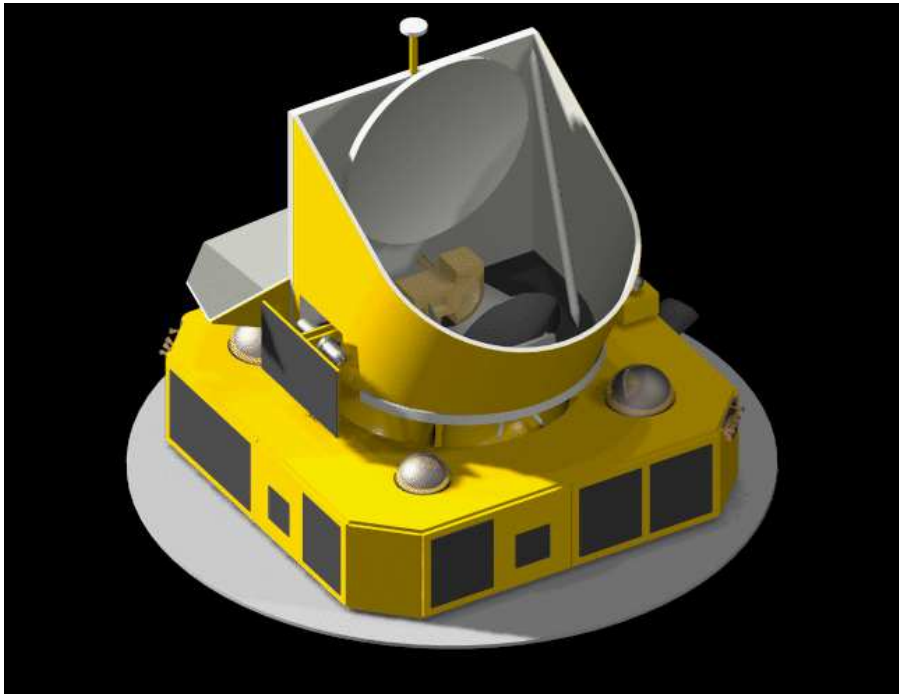,width=12cm}}
\caption{Artist's impression of the Planck Surveyor Satellite
(formally COBRAS/SAMBA)}
\label{fig:COSA-pic}
\end{figure}

\begin{figure}[t]
\centerline{\psfig{file=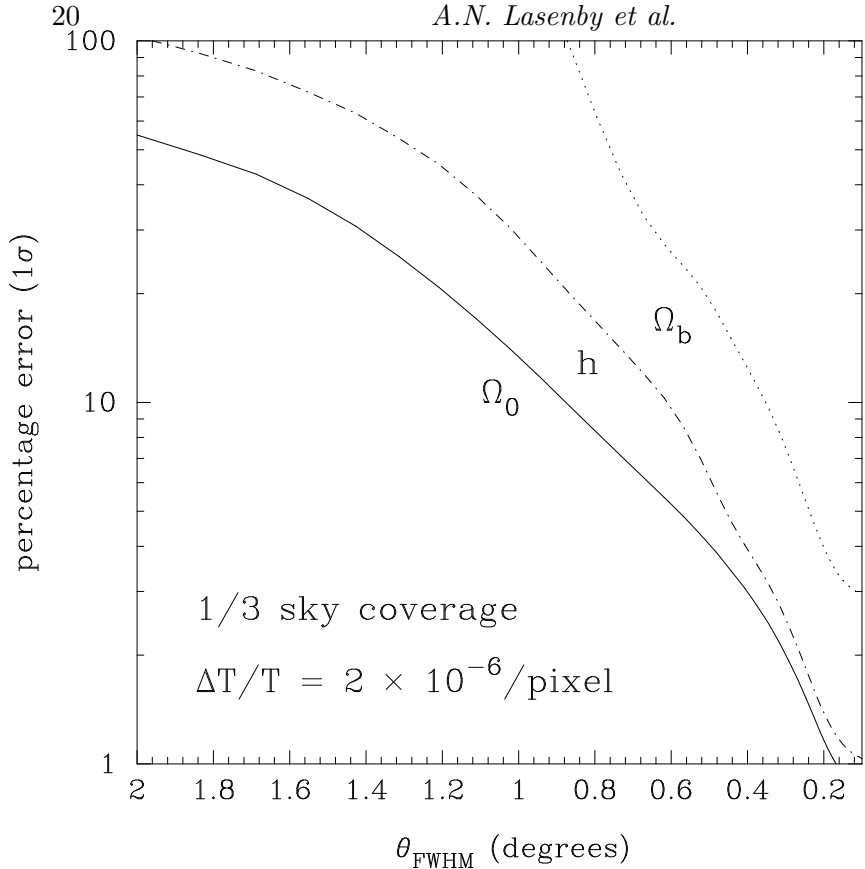,width=13cm}}
\caption{Expected capability of a satellite experiment as a function of
resolution. The percentage error in recovering cosmological parameters from the
CMB power spectrum is shown versus the resolution available. This
figure is taken from Bersanelli \etal \protect\shortcite{bersanelli96}.}
\label{fig:sat-vs-resn}
\end{figure}

Two new satellite experiments to study the CMB have recently been selected as
future missions. These are MAP, or Microwave Anisotropy Probe, which has been
selected by NASA as a Midex mission, for launch in late 2000,
and the Planck Surveyor, which has been selected by ESA as an M3 mission, and will be
launched by 2007. An artist's impression of the MAP
satellite, which has five frequency channels from 30~GHz to
100~GHz, with best resolution 12 arcmin,
is shown in Figure~\ref{fig:MAP-pic}.
An artist's impression of the
Planck Surveyor
satellite, which combines both HEMT and bolometer technology in 10 frequency
channels covering the range 30~GHz to 850~GHz, with best
resolution 4 arcmin, is shown in Figure~\ref{fig:COSA-pic}.
Both these missions are of course of huge importance for CMB
research and cosmology, and even well ahead of launch have sparked
off intense theoretical interest, and many new research programmes
in theoretical CMB astronomy and data analysis. There is no space
here to give an adequate coverage of these missions, or the likely
quality of science which will result. We content ourselves with a
single illustration (Fig.~\ref{fig:sat-vs-resn} --- taken from the
Planck Phase A study), which shows the accuracy with which three
of the main cosmological parameters could be recovered, as a
function of resolution, if 1/3 of the sky was measured to an
accuracy of $\Delta T/T = 2 \times 10^{-6}$ per pixel. (Note a
zero cosmological constant was assumed in producing this figure.)
Obviously such accuracy requires good subtraction of contaminating
foregrounds and point sources, but recent advances in data
analysis, particularly involving application of the maximum
entropy method \cite{planck-mem}, suggest that such accuracy is
feasible.

\subsection{New UK instruments proposed}

In the context of a discussion meeting in the UK it is of interest
to look briefly at two ground-based CMB instruments proposed by UK
groups. Funding is currently being sought for each of these, which
perform complementary observations, of CMB polarization, and
secondary effects in the power spectrum.

\subsubsection{AMI --- the Arc Minute Imager \label{sect:ami}}

This instrument is currently being proposed by MRAO, Cambridge as
a follow-up to the VSA and CAT. It will be a 15 horn
interferometer for CMB structure mapping on angular scales from
0.5 to 5 arcmin. It primary use will be to carry out a survey for
protoclusters via the Sunyaev-Zeldovich (SZ) effect
\cite{sunyaev72}. Two possible candidate high redshift clusters
may already have been identified via this technique
\cite{jones-quasar-pair,richards97}. The number of such clusters
expected is strongly cosmology dependent (e.g. Eke \etal
\shortcite{eke96}) and therefore very useful to measure, and will
provide information complementary to that provided by
high-redshift optical surveys. In particular, a crucial feature of
the SZ effect is that (for the same cluster parameters etc.) the
observed microwave decrement is {\em independent of distance}. It
thus provides a very sensitive indicator of both cosmological
model and cluster gas evolution properties at high redshift.

Further cosmological uses for AMI include detection of other
secondary anisotropies in the power spectrum at high $\ell$'s,
(e.g. the Ostriker-Vishniac effect \cite{ostriker86}), and imaging
of the Kaiser-Stebbins effect in cosmic strings (should a string
exist in the field of view), which as discussed above requires
high angular resolution in order to see the non-Gaussian step-like
discontinuity at the string itself.

\subsubsection{CMBpol --- CMB polarization experiment \label{sect:cmbpol}} 

This experiment, being proposed by a collaboration headed by
Walter Gear of MSSL, is a ground-based instrument for measuring
the CMB polarization power spectrum in the intermediate to small
angular scale range. Specifically the aim is to measure about 10
to 20 square degrees of sky to an accuracy of $\sim 1 \, \mu {\rm
K}$ at a resolution of 3 arcmin over a period of 2 years. The
instrument will use bolometric array and be mounted in
Hawaii. Operation could begin in 2001--2002 if the experiment is
funded.

A key experimental feature of CMB polarization is that the
atmosphere itself is not polarized, and also that the polarization
of a specific spot in the sky can be measured by {\em
difference\/} measurements at that single spot.  Thus the
techniques discussed above in Section~3b for
eliminating atmospheric noise, can be used without any
beam-switching, through the same atmospheric column. This is what
will allow such a sensitive measurement of polarization to be made
from the ground, rather than having to be made from a
satellite. Of course the total intensity component is not
simultaneously available, and only a restricted range of $\ell$'s
will be measured in polarization, due to the finite size of the
sky patch. However, the experiment should provide the first
measurement of the expected sequence of peaks in the polarization
power spectrum (see Fig.~\ref{fig:polzn-spectrum}), and provide
very exciting complementary information to that which will be
available from the VSA, MAP and balloon experiments by that time.

\section{Conclusion}

CMB experiments are already providing significant constraints on
cosmological models, and future experiments will sharpen these up
considerably.  Although full-sky high resolution satellite
experiments like Planck Surveyor will eventually provide
definitive answers for the CMB, the ability of ground-based
experiments and long-duration balloons to go deep on selected
patches, promises to provide very interesting information within
the next two years, followed shortly by the first results from
MAP. Combined with data from large scale structure, supernovae
distances, cluster abundances and other indicators, the next few
years promise to be extremely interesting for cosmology!

\section*{Acknowledgements}

We would like to acknowledge collaboration with several members of
the Cavendish Astrophysics Group, particularly Jo Baker, Sarah
Bridle, Michael Hobson, Michael Jones, Richard Saunders, Paul
Scott and others involved in the CAT group. We would also like to
acknowledge collaborations with several colleagues at IAC
Tenerife, NRAL Jodrell Bank, with Graca Rocha at Kansas and with
George Efstathiou, Ofer Lahav and Matthew Webster at the IoA
Cambridge.

\bibliographystyle{unsrt}

\end{document}